\documentstyle[12pt,a4wide]{article}
\newcommand{\oone}{\hbox{$1\kern-2.5pt\hbox{\rm l}$}}
\newcommand{\ssigma}{\hbox{$\kern2.5pt\vrule height4pt\kern-2.5pt\sigma$}}

\newcommand{\GeV}{{\sl\,GeV}}
\newcommand{\real}{{\sl Re\,}}
\newcommand{\Li}{{\rm Li}_2}
\newcommand\pfrac[2]{\left(\frac{#1}{#2}\right)}
\newcommand{\as}{{\alpha_s}}
\newcommand{\cz}{\chi_{\scriptscriptstyle Z}}
\newcommand{\sxi}{{\sqrt\xi}}

\begin{document}
\thispagestyle{empty}
\begin{flushright}
MZ-TH/97-27\\
hep-ph/9708367\\
August 1997\\
\end{flushright}
\vspace{0.5cm}
\begin{center}
{\Large\bf {\boldmath $O(\as)$} Corrections to Longitudinal}\\[.3cm]
{\Large\bf Spin-Spin Correlations in {\boldmath $e^+e^-\to q\bar q$}}\\[1.3cm]
{\large S.~Groote, J.G.~K\"orner and J.A.~Leyva\footnote{On leave of 
absence from CIF, Bogot\'a, Colombia}}\\[1cm]
Institut f\"ur Physik, Johannes-Gutenberg-Universit\"at,\\[.2cm]
Staudinger Weg 7, D-55099 Mainz, Germany\\
\end{center}
\vspace{1cm}

\begin{abstract}\noindent
We calculate the $O(\as)$ corrections to longitudinal spin-spin 
correlations in $e^+e^-\to q\bar q$. For top quark pair production the 
$O(\as)$ corrections to the longitudinal spin-spin asymmetry amount to 
less than $1\%$ in the $q^2$-range from above $t \bar t$-treshold up to 
$\sqrt{q^2}= 1000\GeV$. In the $e^+e^-\to b\bar b$ case the $O(\as)$ 
corrections reduce the asymmetry value from its $m=0$ value of $-1$ to 
approximately $-0.96$ for $q^2$-values around the $Z$-peak. This reduction 
can be traced to finite anomalous contributions from residual mass effects
which survive the $m\to 0$ limit. We discuss the role of the 
anomalous contributions and the pattern of how they contribute to 
spin-flip and no-flip terms.
\end{abstract}

\newpage\noindent
Recently there has been renewed interest in the role of quark mass effects 
in the production of quarks and gluons in $e^+e^-$ annihilations. Jet 
definition schemes, event shape variables and heavy flavour momentum 
correlations are affected by the presence of quark masses for charm 
and botton quarks even when they are produced at the scale of the 
$Z$-mass~\cite{spin1,spin2,spin3}. A careful investigation of quark mass 
effects in $e^+e^-$ annihilations may even lead to a alternative 
determination of the quark mass values~\cite{spin1,spin2,spin3,spin4}. 
There is obvious interest in quark mass effects for $t\bar t$ production 
where quark mass effects cannot be neglected in the envisaged range of 
energies to be covered by the Next Linear Collider (NLC). In the 
calculation of radiative corrections to quark polarization variables 
residual mass effects change the naive $m=0$ pattern of polarization 
results that results from the absence of spin-flip contributions in the 
$m\to 0$ case~\cite{spin5,spin6,spin7,spin8}.

In this report we calculate the $O(\as)$ 
radiative corrections to longitudinal spin-spin correlations of massive 
quark pairs produced in $e^+e^-$ annihilations. The longitudinal 
polarization of massive quarks affects the shape of the energy spectrum of 
their secondary decay leptons. Thus longitudinal spin-spin correlation 
effects in pair produced quarks and antiquarks will lead to correlation 
effects of the energy spectra of their secondary decay leptons and 
antileptons. As a byproduct of our calculation we discuss the 
$m\to 0$ limit and the role of the $O(\as)$ residual mass effects. We 
delineate how residual mass effects contribute to the various spin-flip 
and no-flip terms in the $m\to 0$ limit.

Let us begin with by defining the differential joint quark-antiquark density 
matrix $d\ssigma=d\sigma_{\lambda_1\lambda_2;\lambda'_1\lambda'_2}$ where 
$\lambda_1$ and $\lambda_2$ denote the helicities of the quark and antiquark, 
respectively. In this paper our main interest is in the longitudinal 
polarization of the quark and antiquark, and in particular, in their 
longitudinal spin-spin correlations. Thus we specify to the diagonal 
case $\lambda_1=\lambda'_1$ and $\lambda_2=\lambda'_2$.

The diagonal part of the differential joint density matrix can be represented 
in terms of its components along the products of the unit matrix and the 
$z$-components of the Pauli matrix $\ssigma_3$ 
($\ssigma_3=\hat{p_1}\vec{\ssigma}$ for the quark and 
$\ssigma_3=\hat{p_2}\vec{\ssigma}$ for the antiquark, 
$\hat{p_i}=\vec{p_i}/|\vec{p_i}|$). One has
\begin{equation}
d\ssigma=\frac 14\bigg(d\sigma\oone\otimes\oone
  +d\sigma^{(\ell_1)}\ssigma_3\otimes\oone
  +d\sigma^{(\ell_2)}\oone\otimes\ssigma_3
  +d\sigma^{(\ell_1\ell_2)}\ssigma_3\otimes\ssigma_3\bigg).
\end{equation}
An alternative but equivalent representation of the longitudinal spin 
contributions can be written down in terms of the longitudinal spin 
components $s_1^{\ell}= 2\lambda_1$ and $s_2^{\ell}= 2\lambda_2$ with 
$s_1^{\ell},s_2^{\ell}=\pm 1$ 
(or $s_1^{\ell},s_2^{\ell}\in\{\uparrow,\downarrow\}$). One has
\begin{equation}\label{eqn1}
d\sigma(s_1^\ell,s_2^\ell)=\frac14\left(d\sigma
  +d\sigma^{(\ell_1)}s_1^\ell+d\sigma^{(\ell_2)}s_2^\ell
  +d\sigma^{(\ell_1\ell_2)}s_1^\ell s_2^\ell\right).
\end{equation}

From $CP$ invariance one knows that $d\sigma$ and $d\sigma^{(\ell_1\ell_2)}$ 
obtain contributions from the parity-even $VV$- and $AA$-current products, 
whereas $d\sigma^{(\ell_1)}$ and $d\sigma^{(\ell_2)}$ are contributed to by 
the parity-odd $V\!A$ and $AV$-current products. The parity-even terms 
$d\sigma$ and $d\sigma^{(\ell_1\ell_2)}$ are $C$-even and thus symmetric 
under $q\leftrightarrow\bar q$ exchange, whereas the parity-odd terms are 
$C$-odd and thus one has 
$d\sigma^{(\ell_1)}(p_1,p_2)=-d\sigma^{(\ell_2)}(p_2,p_1)$ 
for the single-spin dependent contributions.

Eq.~(\ref{eqn1}) is easily inverted. One has
\begin{eqnarray}
d\sigma&=&d\sigma(\uparrow\uparrow)+d\sigma(\uparrow\downarrow) 
  +d\sigma(\downarrow\uparrow)+d\sigma(\downarrow\downarrow)\nonumber\\
d\sigma^{(\ell_1)}&=&d\sigma(\uparrow\uparrow)+d\sigma(\uparrow\downarrow) 
  -d\sigma(\downarrow\uparrow)-d\sigma(\downarrow\downarrow)\nonumber\\
d\sigma^{(\ell_2)}&=&d\sigma(\uparrow\uparrow)-d\sigma(\uparrow\downarrow) 
  +d\sigma(\downarrow\uparrow)-d\sigma(\downarrow\downarrow)\\
d\sigma^{(\ell_1\ell_2)}&=&d\sigma(\uparrow\uparrow)
  -d\sigma(\uparrow\downarrow) 
  -d\sigma(\downarrow\uparrow)+d\sigma(\downarrow\downarrow).\nonumber
\end{eqnarray}

$O(\as)$ radiative corrections to the rate component $d\sigma$ have 
been discussed before~\cite{spin5,spin9} including beam polarization 
effects~\cite{spin6} and beam-event correlation effects~\cite{spin6,spin7}. 
The $O(\as)$ radiative corrections to the longitudinal spin component 
$d\sigma^{(\ell_1)}$ have been recently calculated~\cite{spin5} including 
also beam polarization and beam-event correlation effects~\cite{spin6}. As 
concerns the longitudinal spin-spin correlation component 
$d\sigma^{(\ell_1\ell_2)}$ the $O(\as)$ tree graph contributions have 
been determined in~\cite{spin10}. Here we calculate the $O(\as)$ 
radiative corrections to the fully integrated spin-spin correlation component 
$\sigma^{(\ell_1\ell_2)}$ where we average out beam-event correlation effects.

As before we write the electro-weak cross section 
$e^+e^-\to q(p_1)\bar q(p_2)$ and $e^+e^-\to q(p_1)\bar q(p_2)g(p_3)$ in 
modular form in terms of two building blocks~\cite{spin6}. Thus we write (beam 
polarization effects not included and beam-event correlations averaged out)
\begin{eqnarray}
d\sigma(s_1^{\ell},s_2^{\ell})&=&\frac14\bigg(
  g_{11}(d\sigma_1+d\sigma_1^{(\ell_1\ell_2)}s_1^\ell s_2^\ell) 
  +g_{12}(d\sigma_2+d\sigma_2^{(\ell_1\ell_2)}s_1^\ell s_2^\ell)\nonumber\\&&
  +g_{14}(d\sigma_4^{(\ell_1)}s_1^\ell+d\sigma_4^{(\ell_2)}s_2^\ell)\bigg).
\end{eqnarray}
The index $i=1,2,4$ on the rate components is explained later on.

The first building block $g_{ij}$ ($i,j=1,2,4$) specifies the electro-weak 
model dependence of the $e^+e^-$ cross section. For the present discussion 
we need the components $g_{11}$, $g_{12}$, $g_{14}$, $g_{41}$, $g_{42}$ and 
$g_{44}$. They are given by
\begin{eqnarray}
g_{11}&=&Q_f^2-2Q_fv_ev_f\real\cz+(v_e^2+a_e^2)(v_f^2+a_f^2)|\cz|^2,\nonumber\\
g_{12}&=&Q_f^2-2Q_fv_ev_f\real\cz+(v_e^2+a_e^2)(v_f^2-a_f^2)|\cz|^2,\nonumber\\
g_{14}&=&2Q_fv_ea_f\real\cz-2(v_e^2+a_e^2)v_fa_f|\cz|^2,\nonumber\\
g_{41}&=&2Q_fa_ev_f\real\cz-2v_ea_e(v_f^2+a_f^2)|\cz|^2,\\
g_{42}&=&2Q_fa_ev_f\real\cz-2v_ea_e(v_f^2-a_f^2)|\cz|^2,\nonumber\\
g_{44}&=&-2Q_fa_ea_f\real\cz+4v_ea_ev_fa_f|\cz|^2 \nonumber
\end{eqnarray}
where, in the Standard Model, 
$\chi_Z(q^2)=gM_Z^2q^2/(q^2-M_Z^2+iM_Z\Gamma_Z)^{-1}$, with $M_Z$ and 
$\Gamma_Z$ the mass and width of the $Z^0$ and 
$g=G_F(8\sqrt2\pi\alpha)^{-1}\approx 4.49\cdot 10^{-5}\GeV^{-2}$. $Q_f$ are 
the charges of the final state quarks to which the electro-weak currents 
directly couple; $v_e$ and $a_e$, $v_f$ and $a_f$ are the electro-weak 
vector and axial vector coupling constants. For example, in the 
Weinberg-Salam model, one has $v_e=-1+4\sin^2\theta_W$, $a_e=-1$ for 
leptons,\break $v_f=1-\frac83\sin^2\theta_W$, $a_f=1$ for up-type quarks 
($Q_f=\frac23$), and $v_f=-1+\frac43\sin^2\theta_W$, $a_f=-1$ for down-type 
quarks ($Q_f=-\frac13$). In this paper we use Standard Model couplings
with $\sin^2\theta_W=0.226$.

The second building block is determined by the hadron dynamics, i.e. by the 
current-induced production of a quark-antiquark pair which, in the $O(\as)$ 
case, is followed by gluon emission. In the $O(\as)$ case one also has to 
add the one loop-contribution. We shall work in terms of unpolarized and 
polarized hadron tensor components $H_{U+L}$, $H_{U+L}^{(\ell_1)}$, 
$H_{U+L}^{(\ell_2)}$ and $H_{U+L}^{(\ell_1 \ell_2)}$ where the spin 
decomposition is defined in complete analogy to Eq.~(\ref{eqn1}). In the two 
body case $e^+e^-\to q\bar q$ the unpolarized rate components are given by
\begin{equation}\label{eqn2}
\sigma^i=\frac{\pi\alpha^2v}{3q^4}H^i_{U+L}.
\end{equation}
In the three-body case $e^+e^-\to q\bar qg$ the unpolarized differential 
rate components and the unpolarized hadron tensor components $H_{U+L}^i$ are 
related by
\begin{equation}\label{eqn3}
\frac {d\sigma^i}{dydz}=\frac\alpha{48\pi q^2}H_{U+L}^i(y,z)\quad(i=1,2).
\end{equation}
As kinematic variables we use the two energy-type variables $y=1-2p_1q/q^2$ 
and $z=1-2p_2q/q^2$. The same relations hold for polarized production.

The index $i=1,2$ in Eqs.~(\ref{eqn2}) and~(\ref{eqn3}) specifies the 
current composition in terms of the two parity-even products of the vector 
and the axial vector currents according to (dropping all further indices on 
the hadron tensor)
\begin{equation}
H_{\mu\nu}^1=\frac12(H_{\mu\nu}^{VV}+H_{\mu\nu}^{AA})\qquad
H_{\mu\nu}^2=\frac12(H_{\mu\nu}^{VV}-H_{\mu\nu}^{AA}).
\end{equation}
In the parity-odd case one has
\begin{equation}
H_{\mu\nu}^4=\frac12(H_{\mu\nu}^{VA}+H_{\mu\nu}^{AV}).
\end{equation}
The notation closely follows the one in~\cite{spin6}. Thus the nomenclature 
$(U+L)$ in Eqs.~(\ref{eqn2}) and~(\ref{eqn3}) denotes the total rate ($U$:
unpolarized transverse, $L$: longitudinal) after averaging over the 
relative beam-event orientation.

The generalization of the above cross section expressions to the case where 
one starts with longitudinally polarized beams is straightforward and 
amounts to the replacement
\begin{eqnarray}
g_{14}&\rightarrow&[(1-h^-h^+)g_{14}+(h^--h^+)g_{44}]\\
g_{1i}&\rightarrow&[(1-h^-h^+)g_{1i}+(h^--h^+)g_{4i}]\quad(i=1,2)
\end{eqnarray}
where $h^-$ and $h^+$ ($-1\le h^\pm\le+1$) denote the longitudinal 
polarization of the electron and the positron beam.

Let us begin with by listing the Born term contributions to the various 
polarized and unpolarized two-body hadron tensor components. One has
($\xi=4m_q^2/q^2$, $v=\sqrt{1-\xi}$)
\begin{eqnarray}
H_{U+L}^{1}({\it Born})=(4-\xi)N_C q^2,\qquad&&
H_{U+L}^{2}({\it Born})=3\xi N_C q^2,\nonumber \\
H_{U+L}^{1(\ell_1\ell_2)}({\it Born})=-(4-3\xi)N_Cq^2,\qquad&&
H_{U+L}^{2(\ell_1\ell_2)}({\it Born})=-\xi N_Cq^2,\\
H_{U+L}^{4\,\ell_1 }({\it Born})=4vN_Cq^2,\qquad&&
H_{U+L}^{4\,\ell_2}({\it Born})=-4vN_Cq^2.\nonumber
\end{eqnarray}

The $O(\as)$ spin dependent hadronic tree-body tensor
\begin{equation}
H_{\mu\nu}(p_1,p_2,p_3,s_1,s_2)=\sum_{\mbox{gluon spin}}
\langle q\bar qg|j_\mu|0\rangle\langle 0|j_\mu^\dagger|q\bar qg\rangle
\end{equation}
can easily be calculated from the relevant Feynman diagrams. The 
$(U+L)$-component is then obtained by contraction with the four-transverse 
metric tensor $(-g_{\mu\nu}+q_\mu q_\nu/q^2)$. Finally, the longitudinal 
spin components of the quark and antiquark can be projected out with the 
help of the respective longitudinal spin vectors. They read
\begin{eqnarray}
(s_1^\ell)^\mu&=&\frac{s_1^\ell}\sxi(\sqrt{(1-y)^2-\xi};0,0,1-y)\\
(s_2^\ell)^\mu&=&\frac{s_2^\ell}\sxi(\sqrt{(1-z)^2-\xi};
  (1-z)\sin\theta_{12},0,(1-z)\cos\theta_{12})\nonumber
\end{eqnarray}
with
\begin{equation}
\cos\theta_{12}=\frac{yz+y+z-1+\xi}{\sqrt{(1-y)^2-\xi}\sqrt{(1-z)^2-\xi}}.
\end{equation}
The resulting spin-independent and single-spin dependent components of the 
hadron tensor have been given before (see e.g.~\cite{spin5,spin6,spin9}). 
Here we list the spin-spin dependent piece. One has 
($v_y:=\sqrt{(1-y)^2-\xi}$, $v_z:=\sqrt{(1-z)^2-\xi}$)
\begin{eqnarray}
H_{U+L}^{1(\ell_1\ell_2)}(y,z)\!&=&\!\frac1{v_yv_z}\Bigg[-4(12-10\xi+\xi^2)
  +(1-\xi)(4-3\xi)\xi\left(\frac1{y^2}+\frac1{z^2}\right)\nonumber\\&&
  +(4-3\xi)(8-7\xi)\left(\frac1y+\frac1z\right)+2(12-5\xi)(y+z)\nonumber\\&&
  -2(4-\xi)(y^2+z^2)-(4-3\xi)\xi\left(\frac y{z^2}-\frac{y^2}{z^2}
  +\frac z{y^2}-\frac{z^2}{y^2}\right)\nonumber\\&&
  -2(1-\xi)(2-\xi)(4-3\xi)\frac1{yz}
  -(4-\xi)(6-5\xi)\left(\frac yz+\frac zy\right)\nonumber\\&&
  +2(4-5\xi)\left(\frac{y^2}z+\frac{z^2}y\right)+4\xi yz\Bigg],\\[12pt]
H_{U+L}^{2(\ell_1\ell_2)}(y,z)\!&=&\!\frac\xi{v_yv_z}\Bigg[-4\xi
  +(1-\xi)\xi\left(\frac1{y^2}+\frac1{z^2}\right)
  +(8-7\xi)\left(\frac1y+\frac1z\right)\nonumber\\&&
  -6(y+z)-2(y^2+z^2)-\xi\left(\frac y{z^2}-\frac{y^2}{z^2}+\frac z{y^2}
  -\frac{z^2}{y^2}\right)\nonumber\\&&
  -2(2-\xi)(1-\xi)\frac1{yz}-(6+\xi)\left(\frac yz+\frac zy\right)
  +2\left(\frac{y^2}z+\frac{z^2}y\right)-4yz\Bigg].
\end{eqnarray}

What remains to be done is to perform the phase space integrations and to 
add in the one-loop contributions. In this calculation we perform the 
requisite two-fold phase space integration over the full $(y,z)$ phase space. 
As in~\cite{spin5,spin6} the infrared singularities are regularized by 
introducing a gluon mass. The infrared singularities in the tree graph and 
one-loop contributions cancel and one remains with finite remainders. For 
the sake of completeness we include in our results also the unpolarized 
hadron tensor components which are needed for the normalization of the 
longitudinal spin-spin asymmetry. The $O(\as)$ corrections (tree plus loop) 
read
\begin{eqnarray}
H_{U+L}^1(\as)\!&=&\!N\Bigg[\frac32(4-\xi)(2-\xi)v
  +\frac14(192-104\xi-4\xi^2+3\xi^3)t_3\nonumber\\&&
  -2(4-\xi)\bigg((2-\xi)(t_8-t_9)+2v(t_{10}+2t_{12})\bigg)
  \Bigg],\\[12pt]
H_{U+L}^2(\as)\!&=&\!N\xi\Bigg[\frac32(18-\xi)v
  +\frac34(24-8\xi-\xi^2)t_3\nonumber\\&&
  -6\bigg((2-\xi)(t_8-t_9)+2v(t_{10}+2t_{12})\bigg)\Bigg],\\[12pt]
H_{U+L}^{1(\ell_1\ell_2)}(\as)\!&=&\!N\Bigg[
  (88-78\xi-5\xi^2+3\xi^3)\frac1{2v}
  -40+32\sxi-14\xi+12\xi\sxi+3\xi^2+3\xi^2\sxi\nonumber\\&&
  -\bigg(16-42\xi+31\xi^2-4\xi^3+8(4-3\xi)v^3\bigg)\frac{t_3}{v^2}
  \nonumber\\&&
  +2(4-3\xi)\bigg((2-\xi)(t_8-t_{16})+2v(t_{10}+2t_{12})
  -(4-\xi)(8-3\xi-\xi^2)\frac{t_{13}}{4v^2}\bigg)\nonumber\\&&
  -2(8-10\xi+\xi^2)t_{14}+(32-88\xi+76\xi^2-19\xi^3)\frac{t_{15}}{v^3}
  \Bigg],\\[12pt]
H_{U+L}^{2(\ell_1\ell_2)}(\as)\!&=&\!N\xi\Bigg[
  -(54-65\xi+3\xi^2)\frac1{2v}
  +58-56\sxi-3\xi-3\xi\sxi\nonumber\\&&
  +\bigg(2+\xi-4\xi^2-8v^3\bigg)\frac{t_3}{v^2}
  +2\bigg((2-\xi)(t_8-t_{16})+2v(t_{10}+2t_{12})\bigg)\\&&
  +(96-140\xi+35\xi^2-3\xi^3)\frac{t_{13}}{2v^2}
  -2(10+3\xi)t_{14}+(8-20\xi+13\xi^2)\frac{t_{15}}{v^3}\Bigg]\nonumber
\end{eqnarray}
where we have used an overall normalization factor $N=\as N_CC_Fq^2/4\pi v$. 
The unpolarized hadron tensor components $H_{U+L}^1(\as)$ and 
$H_{U+L}^{2}(\as)$ including the $O(\as)$ rate functions $t_i$ 
($i=3,8,9,10,12$) have been calculated before in~\cite{spin6}. They are 
listed here for completeness. In addition to the rate functions calculated 
in~\cite{spin6} the spin-spin contributions bring in a set of new rate 
functions $t_i$ ($i=13,14,15,16$). The complete set of rate functions needed 
in the present application is given by
\begin{eqnarray}
t_3&=&\ln\left(\frac{1+v}{1-v}\right),\nonumber\\
t_8&=&\ln\left(\frac\xi 4\right)\ln\left(\frac{1+v}{1-v}\right)
  +\Li\left(\frac{2v}{1+v}\right)-\Li\left(-\frac{2v}{1-v}\right)-\pi^2,
  \nonumber\\
t_9&=&2\ln\left(\frac{2(1-\xi)}\sxi\right)
  \ln\left(\frac{1+v}{1-v}\right)
  +2\left(\Li\left(\frac{1+v}2\right)-\Li\left(\frac{1-v}2\right)\right)
  \,+\nonumber\\&&
  +3\left(\Li\left(-\frac{2v}{1-v}\right)
  -\Li\left(\frac{2v}{1+v}\right)\right),\nonumber\\
t_{10}&=&\ln\left(\frac4\xi\right),\quad
t_{12}\ =\ \ln\left(\frac{4(1-\xi)}\xi\right),\quad
t_{13}\ =\ \ln\pfrac{1+v}{2-\sxi},\nonumber\\
t_{14}&=&\ln\pfrac4\xi\ln\pfrac{1+v}{2-\sxi}\nonumber\\&&
  +2\Li\pfrac{2-\sxi}2-2\Li\pfrac\sxi2+\Li\pfrac{1-v}2-\Li\pfrac{1+v}2,
  \nonumber\\
t_{15}&=&\left(\ln\pfrac{1+v}{1-v}+\ln\pfrac\sxi{2-\sxi}\right)^2\nonumber\\&&
  -4\Li\left(\sqrt{\frac{1-v}{1+v}}\right)
  +2\Li\pfrac{2-\sxi}{1+v}+2\Li\pfrac{1-v}{2-\sxi},\nonumber\\
t_{16}&=&\ln\pfrac{1+v}{1-v}\ln\pfrac{4v^4}{\xi(1+v)^2}
  -\Li\pfrac{2v}{(1+v)^2}+\Li\pfrac{-2v}{(1-v)^2}\nonumber\\&&\qquad\qquad
  +\frac12\Li\left(-\frac{(1-v)^2}{(1+v)^2}\right)
  -\frac12\Li\left(-\frac{(1+v)^2}{(1-v)^2}\right).
\end{eqnarray}

We are now in the position to discuss the normalized longitudinal spin-spin 
correlation function $\langle P^{\ell\ell}\rangle$ which is defined as
\begin{equation}\label{eqn4}
\langle P^{\ell\ell}\rangle
  =\frac{\sigma^{(\ell_1\ell_2)}}{\sigma}
  =\frac{\sigma(\uparrow\uparrow)-\sigma(\uparrow\downarrow) 
  -\sigma(\downarrow\uparrow)+\sigma(\downarrow\downarrow)}
  {\sigma(\uparrow\uparrow)+\sigma(\uparrow\downarrow) 
  +\sigma(\downarrow\uparrow)+\sigma(\downarrow\downarrow)}.
\end{equation}
The mean $\langle P^{\ell\ell}\rangle$ is taken with regard to all phase-space 
variables including the beam-event orientation variables.

Before we present our numerical results we briefly pause to discuss the 
$m\to 0$ limit (or equivalently the $q^2\to\infty$ limit) of the various 
hadron tensor components and the longitudinal spin-spin correlation 
functions. In the $m\to 0$ limit the hadron tensor considerably simplifies 
and one has ($H^2_{U+L}\to 0$) 
\begin{eqnarray}
H^1_{U+L}(s_1^\ell,s_2^\ell)
  &=&\frac14\Big(H^1_{U+L}+H^{1(\ell_1 \ell_2)}_{U+L}s_1^\ell s_2^\ell\Big)
  \nonumber\\
  &=&N_Cq^2\left((1-s_1^\ell s_2^\ell)\left(1+\frac{3\as}{4\pi}C_F\right)
  +\left[\frac43\cdot\frac{3\as}{4\pi}C_Fs_1^\ell s_2^\ell\right]\right).
\end{eqnarray}
By hindsight the $O(\as)$ contribution has been split into a normal piece and 
an anomalous piece (indicated by square brackets). The normal piece 
proportional to $(1-s_1^\ell s_2^\ell)$ is defined in terms of the usual 
massless QCD$(m=0)$ no-flip $O(\as)$ result whereas the anomalous piece is 
a residual mass effect which survives the $m\to 0$ limit.

Let us complete our discussion of the $m\to 0$ limit by also listing the 
corresponding $m\to 0$ limit of the longitudinal single-spin hadron tensor
components. One has
\begin{eqnarray}
H^4_{U+L}(s_1^\ell,s_2^\ell)&=&\frac14\Big(H^{4(\ell_1)}_{U+L}s_1^\ell
  +H^{4(\ell_2)}_{U+L}s_2^\ell\Big)\\
  &=&N_Cq^2\left(1+\frac{3\as}{4\pi}C_F
  -\left[\frac23\cdot\frac{3\as}{4\pi}C_F\right]\right)
  (s_1^{\ell}-s_2^{\ell}).
\end{eqnarray}
In Table~\ref{tab1} we list all $m\to 0$ contributions to the various spin 
configurations where again we have split off the $m=0$ no-flip contributions.

\begin{table}
\begin{center}
\begin{tabular}{|c|c|c||c|c|}\hline
spin&\multicolumn{2}{c||}{$VV$}&\multicolumn{2}{c|}{$VA$}\\\cline{2-5}
config.&$m=0$&$m\to 0$&$m=0$&$m\to 0$\\\hline\hline
$(\uparrow\uparrow)$    &$0$&$\frac43=0+[\frac43]$&$0$&$0$\\\hline
$(\uparrow\downarrow)$  &$2$&$\frac23=2-[\frac43]$
  &$2$&$\frac23=2-[\frac43]$\\\hline
$(\downarrow\uparrow)$  &$2$&$\frac23=2-[\frac43]$
  &$-2$&$-\frac23=-2+[\frac43]$\\\hline
$(\downarrow\downarrow)$&$0$&$\frac43=0+[\frac43]$&$0$&$0$\\\hline
\end{tabular}
\caption{$O(\as)$ corrections to specific spin configurations in QCD($m=0$) 
and QCD($m\to 0$). Entries are given in terms of contributions to the hadron 
tensor components $H^{VV}_{U+L}(s_1,s_2)$ and $H^{VA}_{U+L}(s_1,s_2)$ in 
units of $3\as C_FN_Cq^2/4\pi$. Anomalous contributions are shown in square 
brackets\label{tab1}}
\end{center}\end{table}

At threshold the production cross section is 
dominated by the $s$-wave vector current contribution. One thus has 
$H_{U+L}^1(s_1^\ell,s_2^\ell)=H_{U+L}^2(s_1^\ell,s_2^\ell)$ and 
$H_{U+L}^1(\uparrow\uparrow)=\frac12H_{U+L}^1(\uparrow\downarrow)
=\frac12H_{U+L}^1(\downarrow\uparrow)=H_{U+L}^1(\downarrow\downarrow)$. 
From Eq.~(\ref{eqn4}) one then obtains a threshold value of 
$\langle P^{\ell\ell}\rangle=-1/3$.

Let us now present our numerical results. In Fig.~1 we plot the mean 
longitudinal spin-spin asymmetry $\langle P^{\ell\ell}\rangle$ against 
the c.m.\ energy $\sqrt{q^2}$ for top quark pair production. The 
longitudinal spin-spin asymmetry rises from its threshold value of 
$\langle P^{\ell\ell}\rangle=-1/3$ to around 
$\langle P^{\ell\ell}\rangle=-0.9$ at $\sqrt{q^2}=1000\GeV$. The $O(\as)$ 
correction to the asymmetry amounts to less than $1\%$ in the $q^2$-range 
from above $t\bar t$ treshold to $\sqrt{q^2}=1000\GeV$.

In Fig.~2 we present our results on $\langle P^{\ell\ell}\rangle$ for 
bottom quark pair production starting from $b\bar b$ threshold (where 
$\langle P^{\ell\ell}\rangle=-1/3$) up to $\sqrt{q^2}=100\GeV$. For the 
lower $q^2$-values from threshold to about $30\GeV$ the $O(\as)$ corrections 
are quite small. Starting at around $\sqrt{q^2}=30\GeV$ the $O(\as)$ 
correction become larger. The Born term contribution very quickly acquires 
its asymptotic limiting value $\langle P^{\ell\ell}\rangle=-1$ due to the 
fact that the corrections to the leading term are quadratic in the ratio 
$m/\sqrt{q^2}$. Contrary to this the $O(\as)$ curve remains below the 
naive limiting value of $-1$. From the limiting formula
\begin{equation}
\langle P^{\ell\ell}\rangle
  =-\frac{\displaystyle 1+\frac\as\pi-\left[\frac{4\as}{3\pi}\right]}
  {\displaystyle 1+\frac\as\pi}
\end{equation}
one concludes that a large part of the deviation is made up by the 
anomalous contributions. For example, at the position of the $Z$ pole the 
limiting value of the anomalous contribution to $\langle P^{\ell\ell}\rangle$ 
amounts to $\langle P^{\ell\ell}\rangle({\it anom})=0.048$ ($\as(m_Z)=0.118$). 
From the full calculation one finds 
$\langle P^{\ell\ell}\rangle({\it Born})=-0.996$ and
$\langle P^{\ell\ell}\rangle(\as)=-0.964$. Thus the deviation of 
$\langle P^{\ell\ell}\rangle$ from its naive value of 
$\langle P^{\ell\ell}\rangle=-1$ can be seen to arise to a large part from 
the anomalous contribution.\\[0.7cm]
{\bf Note added in proof:} When preparing this manuscript for publication we 
became aware of a preprint on the same subject by M.M.~Tung, J.~Bernab\'eu 
and J.~Pe\~narrocha~\cite{spin11}.\\[0.7cm]
{\bf Acknowledgements:} This work is partially supported by the BMBF, FRG,
under contract No.\ 06MZ865, and by HUCAM, EU, under contract No.\
CHRX-CT94-0579. S.G. acknowledges financial support by the DFG, FRG.
The work of J.A.L.\ is supported by the DAAD, FRG.

\vspace{1cm}
\centerline{\Large\bf Figure Captions}
\vspace{.5cm}
\newcounter{fig}
\begin{list}{\bf\rm Fig.\ \arabic{fig}:}{\usecounter{fig}
\labelwidth1.6cm\leftmargin2.5cm\labelsep.4cm\itemsep0ex plus.2ex}
\item Energy dependence of $O(1)$ and $O(\as)$ mean longitudinal 
  spin-spin correlations $\langle P^{\ell\ell}\rangle$ in 
  $e^+e^-\rightarrow t\bar t(g)$
\item Energy dependence of $O(1)$ and $O(\as)$ mean longitudinal 
  spin-spin correlations $\langle P^{\ell\ell}\rangle$ in 
  $e^+e^-\rightarrow b\bar b(g)$. The vertical line indicates the 
  $b\bar b$-threshold
\end{list}
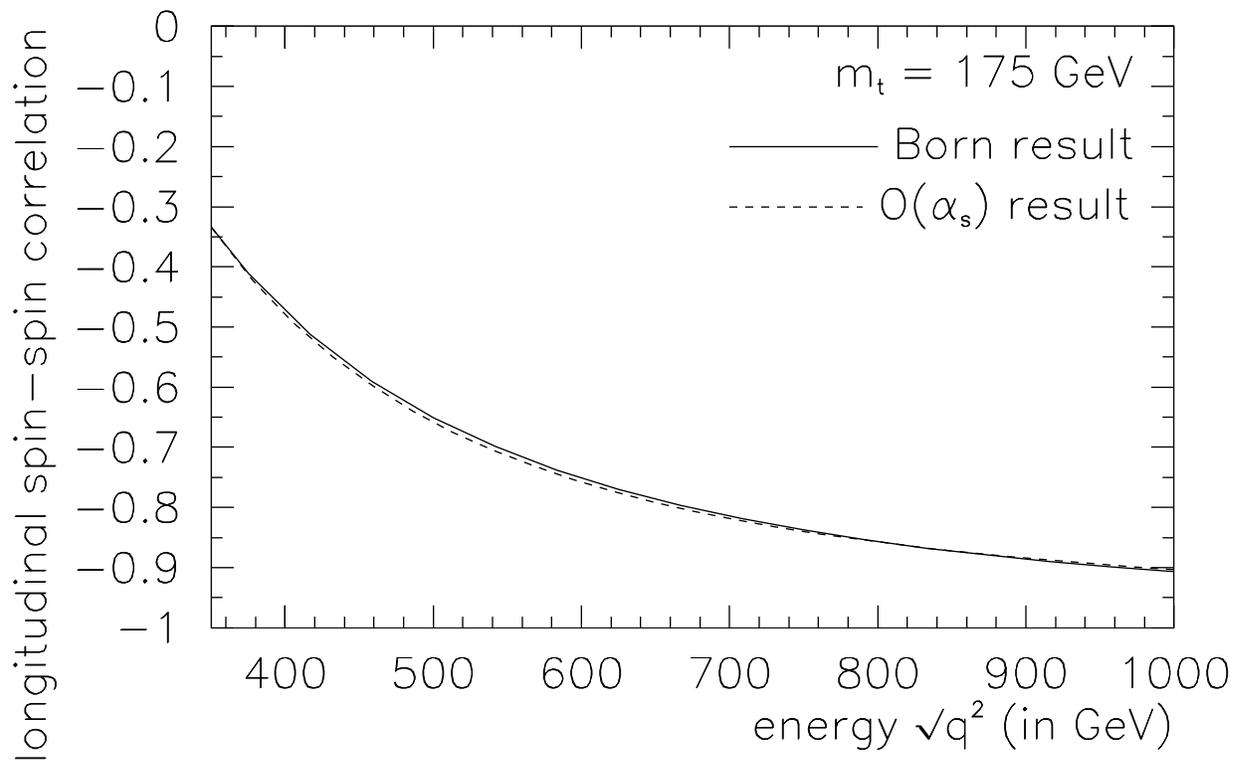
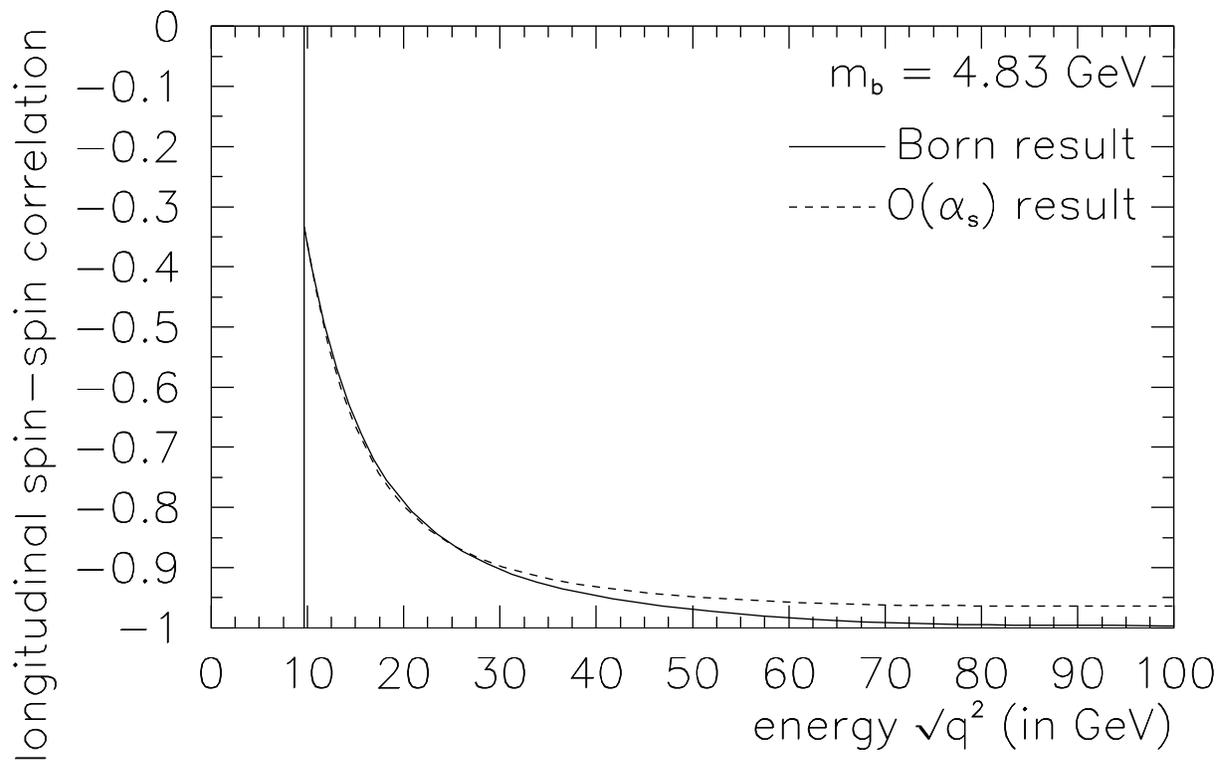
\end{document}